\documentclass[aps,prb,reprint,floatfix,superscriptaddress]{revtex4-2}
\usepackage{graphicx,bm,hyperref}

\begin{document}

\title{Anisotropy at twin interfaces in $RT_{12}$ ($R$=rare earth, $T$=transition metal) magnets} 
\author{Christopher E.\ Patrick}
\email{christopher.patrick@materials.ox.ac.uk}
\affiliation{
Department of Materials, University of Oxford,
Parks Road,
Oxford OX1 3PH, UK}

\date{\today}

\begin{abstract}
RT\textsubscript{12} materials continue to attract attention
due to their potential use as ``rare-earth-lean'' permanent
magnets, but converting their promising intrinsic properties into
practical high performance remains an elusive goal.
Sophisticated experimental characterization
techniques are providing unprecedented insight into the structure
of these materials at the atomistic scale.
Atomistic spin dynamics or micromagnetics simulations could
help unravel the links between these structures
and resultant magnet performance, but require input data
describing the intrinsic magnetic properties.
Here, first-principles calculations based on
density-functional theory are used to determine these
properties for two model interface structures which have
been derived from recently reported high resolution electron 
microscopy
images.
One model structure is a stoichiometric twin formed
by mirroring the RT\textsubscript{12} structure in the 
(101) plane, and the other model structure
is a ``stacking fault'' 
involving the insertion of a RT\textsubscript{4}
plane and a displacement along the [100] axis.
Magnetic moments and crystal
field coefficients have been calculated for the optimized structures.
The interfaces modify the magnetic properties
at the sub-nm scale.
In particular, 
in the R-rich region of the ``stacking fault'',
the local easy axis of magnetization   
rotates by $49^\circ$ from its 
bulk direction, which may lead to reduced coercivity
through the easier nucleation 
of reverse domains. 
\end{abstract}

\maketitle

\section{Introduction}
\label{sec.intro}
The need for permanent magnets with
a reduced rare earth content compared to 
widely-used Nd-Fe-B and Sm-Co
continues to drive research into so-called
``one-twelve'' materials with formula RT\textsubscript{12},
where R is a rare earth element and T is 
a transition metal~\cite{Tozman2021,Gabay2017}.
Alloys of SmFe\textsubscript{12} have 
intrinsic properties which satisfy the key requirements of a permanent magnet,
namely a large volume magnetization, a high Curie temperature,
and a strong uniaxial magnetocrystalline anisotropy~\cite{Hirayama2017,Buschow1991,Schonhobel2019}.
Pure SmFe\textsubscript{12}
is thermodynamically unstable, but stable alloys can be formed
through partial substitution of
Fe with transition metals like Ti and V, or of Sm with
elements like Zr, Gd or Y~\cite{Makuta2025,Xu2024,Tozman2022,Harashima2018,Kuno2016,
Fukazawa20222,Kobayashi2023,Matsumoto2020,Harashima20152}.
Optimizing the composition requires considering both 
performance and cost.
For instance, 
Co is an excellent substitute for Fe
from the point of view of magnetic properties, 
but suffers from similar supply
issues as the rare earths themselves~\cite{Tkaczyk2018}.

\begin{figure*}
    \includegraphics[width=1.8\columnwidth]{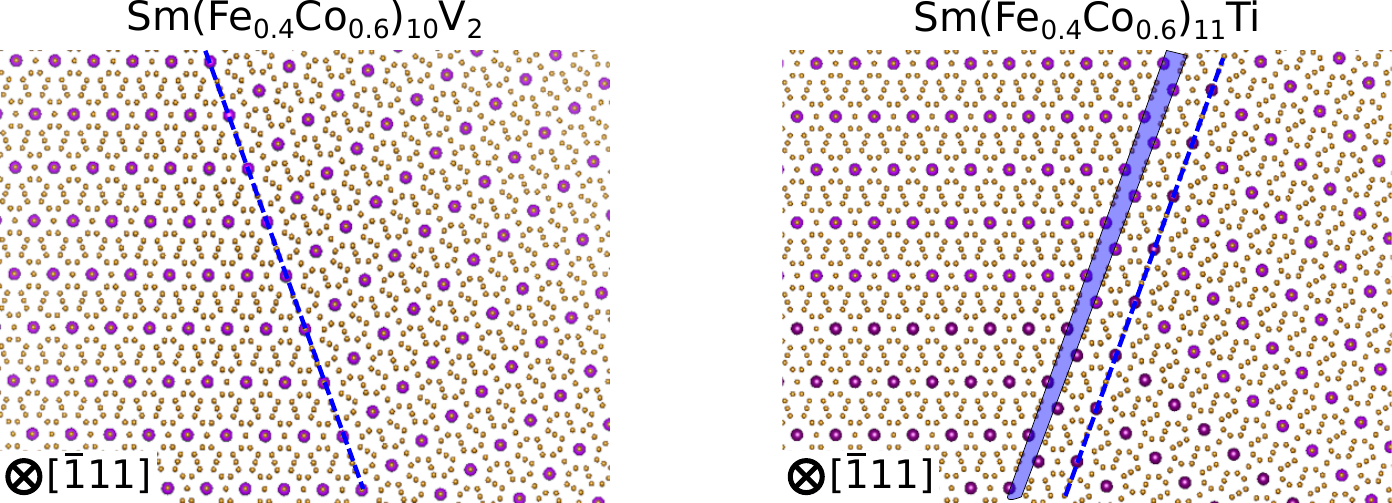}
    \caption{Representations of the two atomistic
    models proposed in the current work which reproduce 
    the HAADF-STEM images 
    shown in Fig. 3 of Ref.~\cite{Tozman2025}.  
    The structures are viewed
    along the $[\bar{1}11]$ direction, which lies in the $(101)$
    twin plane.  R and T atoms are represented as large pink or
    small yellow circles, respectively. 
    Schematics of these structures are shown in Fig.~\ref{fig.fig_construction}.
    \label{fig.fig_models}
    }
\end{figure*}

Aside from cost, the outstanding challenge surrounding one-twelve materials is 
how to convert promising intrinsic behavior into the
key extrinsic properties of remanence
and coercivity~\cite{Palanisamy2020,Hadjipanayis2020,SepehriAmin2020,
Zhang2021,Gabay2022,Bolyachkin2022}.
For these properties, the 
microstructure of the materials is crucial, which creates
the need to characterize and understand the role of secondary
phases and defects.
There has been particular interest
in the presence of twins in one-twelve samples, with boundary regions proposed
to act as nucleation sites for magnetization reversal~\cite{Polin2025,Zhang2022,
Ener2021,Dirba20192}.
Recently, high-resolution high-angle annular dark-field images 
obtained from scanning tunnelling electron 
microscopy (HAADF-STEM) showed twin boundaries 
at an unprecedented level of detail~\cite{Tozman2025}.
In that work,
two markedly different HAADF-STEM images were observed
depending on the composition of the RT\textsubscript{12} alloy.
Atomistic models of these interfaces (as proposed in the
current work) are shown in Fig.~\ref{fig.fig_models}.
The original images appear as Fig.~3 in Ref.~\cite{Tozman2025}.

This ability to visualize the structure of interfaces
at the atomistic scale raises two key questions: (i) how
does the modified atomistic structure affect 
the magnetic properties of atoms in this region, and (ii) what
impact do these modified magnetic properties 
have on the material's macroscopic performance as 
a permanent magnet?
Considering the first question, the local environment (crystal field) 
of R atoms is responsible for their large magnetocrystalline
anisotropy~\cite{Buschow1977}, so any significant perturbation
to the crystal field should affect the anisotropy.
For instance, calculations of the crystal
field of a
Nd\textsubscript{2}Fe\textsubscript{14}B slab found
that Nd atoms at the (001) surface had easy-plane rather than easy-axis
anisotropy~\cite{Moriya2009}.
We may also 
expect modification of the magnetic moments of
the atoms and of the exchange coupling between them, as has also been
shown for Nd-Fe-B interfaces~\cite{Umetsu2016}.

Considering the second question regarding the impact on permanent magnet performance, 
inhomogeneities in the  anisotropy form the basis
of Kronm\"uller's theory of why the (extrinsic) 
coercive field is only a fraction of the 
(intrinsic) anisotropy field,
due to the smaller nucleation fields of reverse domains in 
regions of low anisotropy~\cite{Kronmuller1987}.
Analytical calculations also show how inhomogeneous anisotropy 
leads to pinning of domain walls, which is the relevant
coercivity mechanism for Sm-Co magnets~\cite{Kronmuller2002}.
Advances in the computational modeling
of micromagnetics (MM)~\cite{Fischbacher2018} and atomistic spin 
dynamics (ASD)~\cite{Evans2018} have enabled numerical simulations 
at a higher level of complexity than possible with analytical calculations, 
to gain material-specific insight into magnetization reversal and coercivity.
Some examples are calculating domain wall widths in bulk 
Nd\textsubscript{2}Fe\textsubscript{14}B~\cite{Nishino2017,Gong2019},
domain wall structures in Sm\textsubscript{2}Co\textsubscript{17}~\cite{SepehriAmin2017}
and nucleation/coercive fields in various materials, including 
RT\textsubscript{12} magnets~\cite{Bolyachkin2024,Ener2021,Westmoreland2020,Fischbacher2017}.

Both MM and ASD allow the researcher
to specify a material's microscopic properties and then
carry out numerical ``measurements'' of macroscopic magnetic performance.
The ability of such simulations to obtain realistic outputs
is determined by the accuracy of the equations which make up the model,
and by the quality of the input data.
The required input data are
the intrinsic properties of
magnetization, exchange and anisotropy, resolved
at the atomistic/sub-micron level for ASD/MM respectively.
When simulating a material composed of well-characterized phases,
these intrinsic properties might be derived from experimental measurements
on single crystals of the relevant phases.
However, the interfaces regions shown in 
Fig.~\ref{fig.fig_models} are not crystalline, and
the environments of the 
atoms in the sub-nm interfacial regions do not resemble
bulk phases.
Determining the magnetic properties of these specific 
atoms is therefore a formidable experimental challenge.

First principles calculations provide an attractive alternative
method of obtaining the intrinsic properties of novel structures.
The sub-nm lengthscale is ideally suited to calculations
based on (spin) density-functional theory (DFT)~\cite{Hohenberg1964,Kohn1965}.
Solving the fundamental equations of DFT, including using the Hellman-Feynman
theorem to obtain forces on atoms, allows the first-principles calculation
of structural parameters (e.g.\ bond lengths), and properties deriving
from electrons such as magnetic moments and the crystal field.
Crucially, DFT does not require prior empirical knowledge of the system
under study, only the approximate positions and types of the atoms involved.
In this work, this information is deduced from the HAADF-STM images.
The principal complication associated with DFT calculations on 
rare earth magnets is modeling of exchange and 
correlation,
with the most widely-used approaches
based on the local density or generalized gradient 
approximations struggling
to describe the lanthanide-4$f$ electrons~\cite{Richter1998}.
However, a number of methods have been developed to overcome
this issue and are in active use, such as the open-core approach, DFT+$U$,
the self-interaction correction and dynamical mean-field
theory~\cite{Brooks19912,
Fukazawa2022,Herper2023,Yoshioka2022,Larson2004,Patrick20182,Delange2017}.
In this work, the ``yttrium-analog'' (Y-analog) approach is employed,
where all calculations are performed with Y as the rare earth element~\cite{Patrick20192}.
The justification for this approach is that 
it is the valence electrons
which have the greatest effect on the transition metal magnetism
and the crystal field~\cite{Bouaziz2023}.
Since Y$^{3+}$ and Sm$^{3+}$ have the same $s^2d$ 
valence electronic structure, Y is a convenient analog,
free of 4$f$ electrons, to investigate these properties.
The Y-analog approach was inspired by experimental literature,
where studies on materials such as Y\textsubscript{2}Fe\textsubscript{14}B~\cite{Hirosawa1986},
YCo\textsubscript{5}~\cite{Ermolenko1979} and YFe\textsubscript{11}Ti~\cite{Moze1988} provided valuable insight
into their lanthanide-based counterparts.

The purpose of this study is to use first-principles calculations to obtain the
magnetic properties of atoms in the twin boundary interface regions recently 
observed in one-twelve materials,
thus addressing the first key question posed above.
The HAADF-STEM images from Ref~\cite{Tozman2025} 
are used to derive atomistic models of the interfaces.
Applying DFT to these models yields optimized structures and local magnetic moments.
The crystal field at each R site in the interface models is then calculated
within the Y-analog approach.
Crystal field theory allows the single ion anisotropy energy for Sm-based
magnets to be deduced, giving local anisotropy energies and magnetization directions.
The calculations show that there are changes to the magnetic
moments and crystal field only within
a few \AA \ of the interface.
However, this change in crystal field is substantial for 
one of the structures,  resulting in  a rotation of the
easy direction by $49^\circ$ away
from the $c$-axis.

The rest of this manuscript is ordered as follows.
Section~\ref{sec.methods} describes the construction of 
the interface models in Fig.~\ref{fig.fig_models} and
their relation to the RT\textsubscript{12} structure.
Section~\ref{sec.methods} also describes the
embedding of these structures
in periodically-repeating simulation cells 
and gives the methodological details
of the DFT calculations, including an overview of the 
method of calculating the crystal field coefficients.
Section~\ref{sec.results} presents the results
of the calculations, including radial distribution
functions of the relaxed structures, magnetic moments
and crystal field coefficients.
The crystal field coefficients are used to construct
the single-ion anisotropy energy in a classical approximation.
Finally, Section~\ref{sec.conclusions} presents conclusions
and identifies areas for future study.

\section{Methodology}
\label{sec.methods}
\subsection{Models of extended interface structures}
\begin{figure}
    \includegraphics[width=0.8\columnwidth]{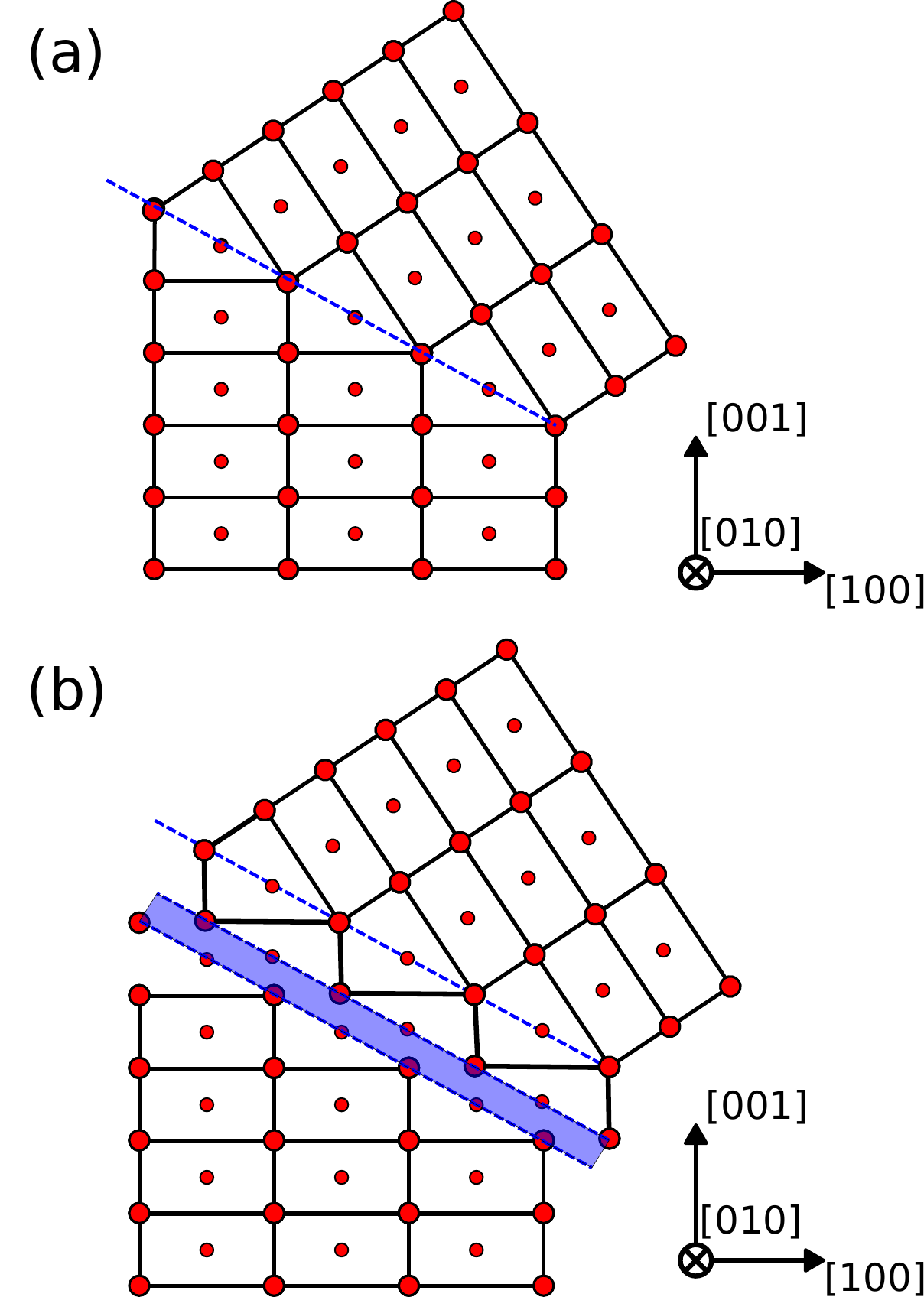}
    \caption{
    The same structures as shown in Fig.~\ref{fig.fig_models}
    of (a) Sm(Fe\textsubscript{0.4}Co\textsubscript{0.6})\textsubscript{10}V\textsubscript{2} and (b) Sm(Fe\textsubscript{0.4}Co\textsubscript{0.6})\textsubscript{11}Ti, viewed along the $[010]$
    direction, with only R sites shown; the larger and smaller
    circles correspond to the sites at the corners and center
    of the bct cell, respectively.  The $(101)$ twin plane
    is shown as the blue dashed line.  Panel (b) also
    shows as a blue shaded rectangle the extended defect formed
    by the cleaving, displacement and insertion of an
    RT\textsubscript{4} plane.  Note
    that the axes refer to the lowest region
    in the figures.
    \label{fig.fig_construction}
    }
\end{figure}

Figure 3 of Ref.~\cite{Tozman2025} reported HAADF-STEM images
of two different structures, both formed from twinning along
\{101\} planes, observed for 
Sm(Fe\textsubscript{0.4}Co\textsubscript{0.6})\textsubscript{10}V\textsubscript{2}
and 
Sm(Fe\textsubscript{0.4}Co\textsubscript{0.6})\textsubscript{11}Ti.
Figure~1 of the current work 
proposes atomistic structures of the interfaces,
based on these images.
The orientations in Fig.~\ref{fig.fig_models} 
have been chosen
to match the images in Ref.~\cite{Tozman2025}, but
Fig.~\ref{fig.fig_construction} gives
a schematic view of the same structures 
which better illustrates how the models were constructed.
RT\textsubscript{12} has a body-centered tetragonal (bct) structure,
with $a > c$.
The rare-earth atoms occupy the $(0,0,0)$ and
$(\frac{1}{2},\frac{1}{2},\frac{1}{2})$ 
positions, and for simplicity we only show
these positions in Fig.~\ref{fig.fig_construction}.
The (101) plane which bisects the conventional cell
contains R and T atoms at a ratio of RT\textsubscript{4},
with the T atoms equally distributed on $8i$ 
and $8f$ Wyckoff sites.
The Sm(Fe\textsubscript{0.4}Co\textsubscript{0.6})\textsubscript{10}V\textsubscript{2}
structure [Fig.~\ref{fig.fig_construction}(a)] is formed simply
by reflecting the crystal
in this plane.
This operation does not affect the 1:12 stoichiometry.
The more complicated 
Sm(Fe\textsubscript{0.4}Co\textsubscript{0.6})\textsubscript{11}Ti
structure 
also has this reflection.
In addition, there is an extended defect which, as shown
in Fig.~\ref{fig.fig_construction}(b), 
is formed by cleaving the crystal along the (101)
plane, and introducing a displacement
between the cleaved sections of $a/2$ in the 
$[100]$ direction.
In addition, an RT\textsubscript{4} plane of atoms 
is introduced into the created gap.
These additional atoms result in
a local stoichiometry which is R-rich compared to the
RT\textsubscript{12} bulk.
Note that the same defect can be described equivalently  as a 
removal of all T atoms lying between a pair
of adjacent (101) planes followed by the introduction of
a displacement
of $a/2$ along $[\bar{1}00]$ between the regions.
In the Sm(Fe\textsubscript{0.4}Co\textsubscript{0.6})\textsubscript{11}Ti
structure, this defect is followed by a reflection 
in the next (101) plane [Fig.~\ref{fig.fig_construction}(b)].
These models provide an excellent
account of the experimentally-observed images in Fig.~3 of Ref.~\cite{Tozman2025}.

\subsection{Periodic models for first-principles calculations}

\begin{figure}
    \includegraphics[width=\columnwidth]{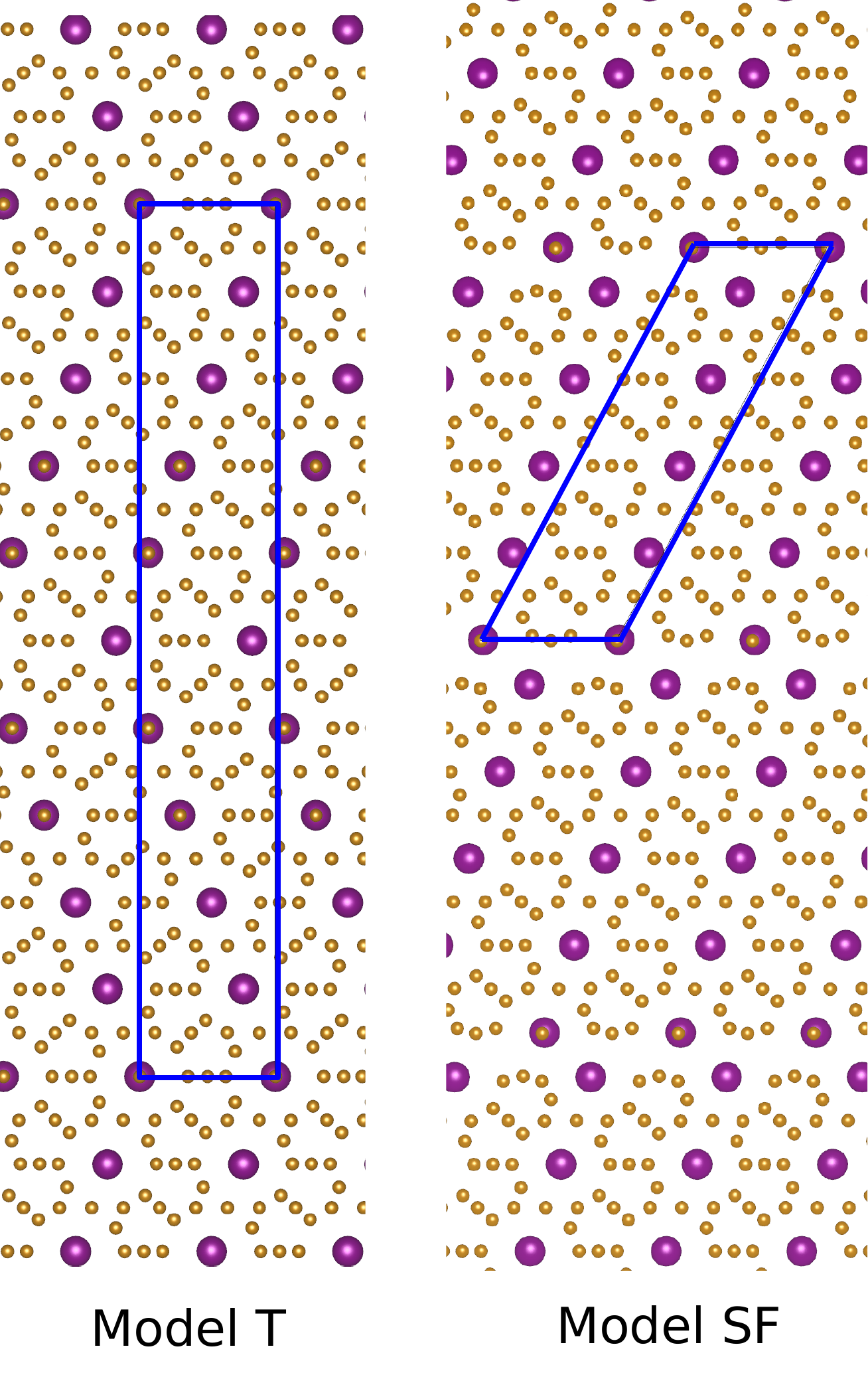}
    \caption{The two periodically-repeating models used in this work.  
    The models are viewed along $[\bar{1}11]$, which coincides with the direction
    of one of the cell vectors.  The boundary of the
    unit cell in the perpendicular plane is shown in blue.  These models were structurally relaxed as described in the text. 
    Note that the R atoms do not all lie in the plane
    of the page.  %
The perpendicular distance between repeated interface planes is
20.5~\AA \ for Model T and 18.5~\AA \ for model SF.
    \label{fig.fig_atomistic}
    }
\end{figure}

The extended interface structures shown in 
Figs.~\ref{fig.fig_models}~and~\ref{fig.fig_construction} 
break translational symmetry.
However, many implementations 
of first-principles electronic structure
theories assume periodically-repeating simulation cells,
which allow efficient and well-converged solutions 
of the numerical equations
through the use of Bloch's theorem and plane wave basis sets~\cite{Ihm1979}.
It is therefore desirable to 
construct models which 
reproduce the relevant atomic environments of Fig.~\ref{fig.fig_models},
but also have periodically-repeating simulation cells.
These models are shown in Fig.~\ref{fig.fig_atomistic}, 
and are labeled as ``twin'' (T) 
or ``stacking fault'' (SF)~\footnote{It should be noted
that ``stacking fault'' is a convenient label which
captures the offset columns of RE atoms, but the structure
is not a conventional stacking fault due to the insertion
of extra planes of atoms.}.
Model T reproduces the stochiometric (101) twinning seen
in both interface structures (blue lines in 
Figs.~\ref{fig.fig_models}~and~\ref{fig.fig_construction}).
Model SF reproduces the motif seen in 
Sm(Fe\textsubscript{0.4}Co\textsubscript{0.6})\textsubscript{11}Ti
structure, whereby
an offset is introduced between adjacent (101) planes
and the local stoichiometry becomes R-rich (shaded 
region of Figs.~\ref{fig.fig_models}~and~\ref{fig.fig_construction}).
These models were constructed firstly by noting that 
the translational symmetry parallel to the (101) plane 
is not broken by the interface, so for both models, one can
choose two of the three simulation cell vectors 
to also lie in this plane, with minimal, equal
lengths of $\sqrt{(2a^2+c^2)}/2$.
The third simulation cell vector is different for the two
models.
For Model T, it is perpendicular to the (101) plane,
and translational symmetry is achieved by having two
twin planes per simulation cell, located at the center
and the edges of the blue rectangle in 
Fig.~\ref{fig.fig_atomistic}.
For Model SF, the third vector is not perpendicular to
the (101) plane, but instead chosen to 
reproduce the stacking fault
with pairs of R atoms at the cell edge 
located at a distance $a/2$ from each other.
The remaining choice is how large to make the simulation cell,
i.e.\ how many RT\textsubscript{4} planes should be included.
A larger number better represents 
the true, aperiodic interface, but increases the computational
expense.
The models in Fig.~\ref{fig.fig_atomistic} were chosen 
to have similar distances ($\sim$20~\AA) 
between repeated interface planes.
This distance is justified by the results presented later
in the manuscript, where it is shown that 
the interface's influence (e.g.\ on local magnetic moments
or the anisotropy) has a much shorter range.
The cell vectors and atomic positions are supplied as
Supplemental Information~\cite{suppinfo}.

\subsection{The Y-analog approach, and choice of T}

As noted in Section~\ref{sec.intro}, a huge variety of
RT\textsubscript{12} alloys can be formed through substitution
of R and/or T.
All of the calculations presented in this work are carried
out on a single composition, which is YFe\textsubscript{12}.
Using Y as the rare earth element is the central idea
of the Y-analog approach~\cite{Patrick20192}.
Its justification is that it is the valence electronic
structure --- the $6s^25d$ electrons for the lanthanides
($5s^24d$ for Y) --- which hybridizes with transition metal
electrons,
setting up the (antiferromagnetic) exchange coupling
and the crystal field~\cite{Patrick20192,Brooks19912}.
In this picture, the lanthanide 4$f$ shell is anchored to
the valence electronic structure through 
the onsite 4$f$-5$d$ interaction, but the 4$f$ electrons
do not themselves influence the electronic structure.
This is an approximation, and calculations that take 
the 4$f$ electrons into account do show differences
with the Y-analog model, for instance in calculating
variations in 
Curie temperatures~\cite{Patrick20182} or higher order
crystal field coefficients~\cite{Pourovskii2020}.
However, for the lower order crystal field coefficients
(which tend to dominate magnetic behavior at working 
temperatures), and to obtain a general picture of magnetic properties, 
the Y-analog model is an attractive choice.
It provides a similar accuracy to the open core 
method~\cite{Patrick20192}, but circumvents any
complications from simulating lanthanide elements in DFT.
Furthermore, the method is numerically robust
thanks to the use of
plane wave basis sets.

For the transition metal, this work focuses exclusively
on T=Fe, and does not take into account any specific effects
of the alloying elements.
Energy-dispersive spectroscopy (STEM-EDS) 
can be used to distinguish between R and T sites in the
HAADF-STEM images in Ref.~\cite{Tozman2025},  but not to
identify individual Fe/Co/Ti/V sites.
Calculations on bulk RT\textsubscript{12} have
been used to identify preferred sites of low
concentrations of dopants, 
e.g.\ Refs.~\cite{Harashima20152,Matsumoto2020},
and in principle a similar analysis could be carried out
for the interface models in Fig.~\ref{fig.fig_atomistic}.
However, a comprehensive exploration of the 
configuration space of the (Fe,Co,Ti/V) alloy, particularly
one which is free from the finite size effects associated
with periodically-repeating simulation cells, would be 
a significant computational undertaking and beyond 
the scope of the current investigation.
Such an exploration is therefore reserved for future
work, which should also aim to elucidate the relationship
between 
a sample's composition and the likelihood that it will form a
particular interface structure.

Whilst focusing on a single T element may seem like a limitation,
recent work has highlighted
the importance of studying the crystal field of
the RFe\textsubscript{12} host.
Ref.~\cite{Patrick2024} demonstrated how the modification to
the crystal field due to Ti substitutions in bulk 
RFe\textsubscript{12} could be well described by a point
charge model (PCM)~\cite{Patrick2024}.
Test calculations associated with that work showed how Co
substitutions did not strongly affect the crystal field,
which is expected from the PCM due to the similar atomic numbers
of Co and Fe.
However, for Ti or V substitution, the effects can be 
significant.
Taking the crystal field of the unperturbed 
RFe\textsubscript{12} host calculated in the current work,
and applying the methodology of
Ref.~\cite{Patrick2024}, will allow the single-ion
anisotropy to be calculated for arbitrary distributions of 
substituting atoms.
As noted in Section~\ref{sec.intro}, these anisotropy energies
are essential input data for larger-scale ASD or MM simulations.

\subsection{Model construction and relaxation}
\label{sec.relaxation}
The models in Fig.~\ref{fig.fig_atomistic} were constructed 
using RT\textsubscript{12} lattice constants
of $a$ = 8.497~\AA \ and $c$ = 4.687~\AA, which were 
previously calculated using DFT for SmFe\textsubscript{12}
in Ref.~\cite{Harashima2015}.
The simulation cell vectors were fixed according to these values
and the positions of all atoms in the cell were relaxed 
until the force on each
atom was below 0.05~eV\AA$^{-1}$.
The same approach was used for bulk calculations which were also carried
out for comparison.
The decision to fix lattice vectors whilst allowing atomic positions
to relax was taken in order to match as closely possible previous
uses of the Y-analog model, and ensure that there were no net forces
on atoms in the unit cell (although an overall strain will still be
present due to the differing lattice constants of YFe\textsubscript{12}
and SmFe\textsubscript{12}).
The calculations used the local-density 
approximation to exchange and correlation~\cite{Perdew1992}, 
and treated the core-valence interaction using the projector-augmented wave (PAW)
method~\cite{Blochl1994}, as implemented the \texttt{GPAW} code~\cite{Enkovaara2010}.
The electronic wavefunctions were expanded in a plane-wave
basis set up to a maximum cutoff energy of 800~eV, and
reciprocal space was sampled using non-shifted grids
of size $4\times4\times1$ for Model T, $4\times4\times2$
for Model SF and $4\times4\times6$ for bulk.
The electronic states were occupied according to
the Fermi-Dirac distribution with width 0.01~eV.

\subsection{Crystal field calculations}

Crystal field coefficients were calculated using the 
methodology described in Ref.~\cite{Patrick20192}.
In this approach,  the DFT Kohn-Sham potential $V(\mathbf{r})$
(which includes both electrostatic interactions and
exchange and correlation~\cite{Kohn1965}) is used to calculate
the crystal field coefficients $B_{lm}$, through the equation
\begin{eqnarray}
    B_{lm} = \left(\frac{2l+1}{4\pi}\right)^\frac{1}{2}
    \int_0^{r_c} r^2 n^0_{4f}(r) V_{lm}(r) dr \nonumber
\end{eqnarray}
In this equation, $n^0_{4f}(r)$ is the normalized 4$f$ electron
charge density, which is obtained in a separate calculation
on bulk SmFe\textsubscript{12} using the self-interaction 
correction~\cite{Lueders2005}.
$V_{lm}(r)$ is obtained from an angular decomposition
of the Kohn-Sham potential using the complex spherical
harmonics, $V(\mathbf{r}) = \sum_{lm}V_{lm}(r)
Y_{lm}(\theta,\phi)$.
Ref.~\cite{Patrick20192} describes the conversion of the Kohn-Sham potential from a plane-wave representation 
to that on a radial grid, including the appropriate PAW corrections.
The radial limit $r_c$ of the integrals was chosen to 
be 1.65~\AA, similar to Ref.~\cite{Patrick20192}.
Convergence checks were carried out to ensure
that the computational parameters described in 
Section~\ref{sec.relaxation} were also sufficiently accurate to calculate
crystal field coefficients.
Also, since the Kohn-Sham potential depends on spin through
the exchange and correlation terms, the method
generates separate crystal field coefficients for
spin up and spin down electrons.
Here, the differences between the two sets are small,
and therefore the presented results are averages over
the spin directions.
Finally, the calculated coefficients $B_{lm}$ are related
to the conventional representation through  
$A_{lm} \langle r^l\rangle = \alpha_{lm} B_{lm}$,
where the prefactors $\alpha_{lm}$ are tabulated in 
Ref.~\cite{Newman1989}.

\section{Results and Discussion}
\label{sec.results}

\subsection{Structural relaxation}
\begin{figure}
    \includegraphics[width=\columnwidth]{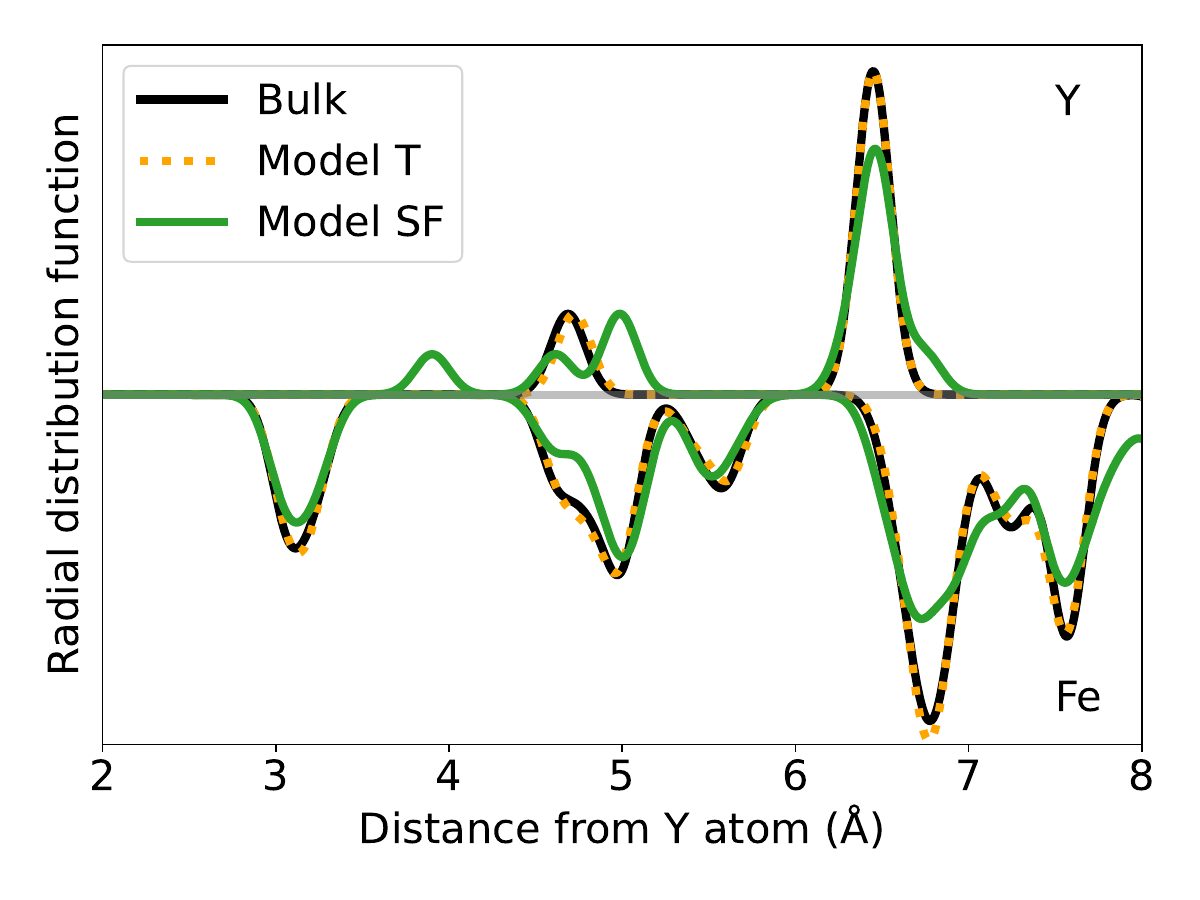}
    \caption{  RDFs for the Y atoms located directly at the 
    interface in Model T or Model SF, and for an Y atom in
    bulk YFe\textsubscript{12}.  The positive or negative scale 
    describes distances to Y or Fe atoms, respectively.
    These RDFs correspond to 
    structures which were relaxed as described in 
    Section~\ref{sec.relaxation}.
    \label{fig.fig_RDF}}
\end{figure}

Figure~\ref{fig.fig_atomistic} shows the
structurally-relaxed interface models.
The only atoms whose positions change noticeably during
the structural relaxation
are those located directly at the interface
in Model SF, in the adjacent YFe\textsubscript{4} planes.
As can be seen by comparing with any of the other 
YFe\textsubscript{4} planes in Fig.~\ref{fig.fig_atomistic},
these Fe atoms are displaced from the (101) plane.
Further analysis of the structure 
can be carried out using radial distribution functions (RDFs). 
Figure~\ref{fig.fig_RDF} shows the RDFs 
for Y atoms located directly at the interface
for each model, compared to the RDF for the bulk structure.
The bulk and Model T RDFs are indistinguishable from each
other.
This emphasizes that the mirror operation shown in 
Fig.~\ref{fig.fig_construction}(a) affects the orientation
of the symmetry axis, but does not introduce
new atoms.
However, the RDFs calculated for Model SF show that the 
local environment of Y atoms at the interface is Y-rich
compared to bulk.
In the unrelaxed structure, one of the two Y atoms
which are located at a distance of 4.69~\AA \ ($=c$) in bulk
is brought closer,
to a distance of 4.25~\AA \ ($=a/2$).
Furthermore, two new Y atoms are introduced 
at a distance of 4.85~\AA \ ($=\sqrt{(a^2+c^2)}/2$).
After relaxation, the nearest Y--Y distance 
reduces from 4.25~\AA \ to 3.90~\AA, 
whilst the distance to the two new Y atoms
increases from 4.85~\AA \ to 4.99~\AA.
Therefore, a Y atom at the interface in Model SF is surrounded
by four other Y atoms within a radius of 5~\AA, compared to bulk
or Model T where there are only two.
Furthermore, in Model SF there are two fewer Fe atoms in the first coordination shell (18)
compared to bulk or Model T (20).

\subsection{Magnetic moments}
\begin{figure}
    \includegraphics[width=\columnwidth]{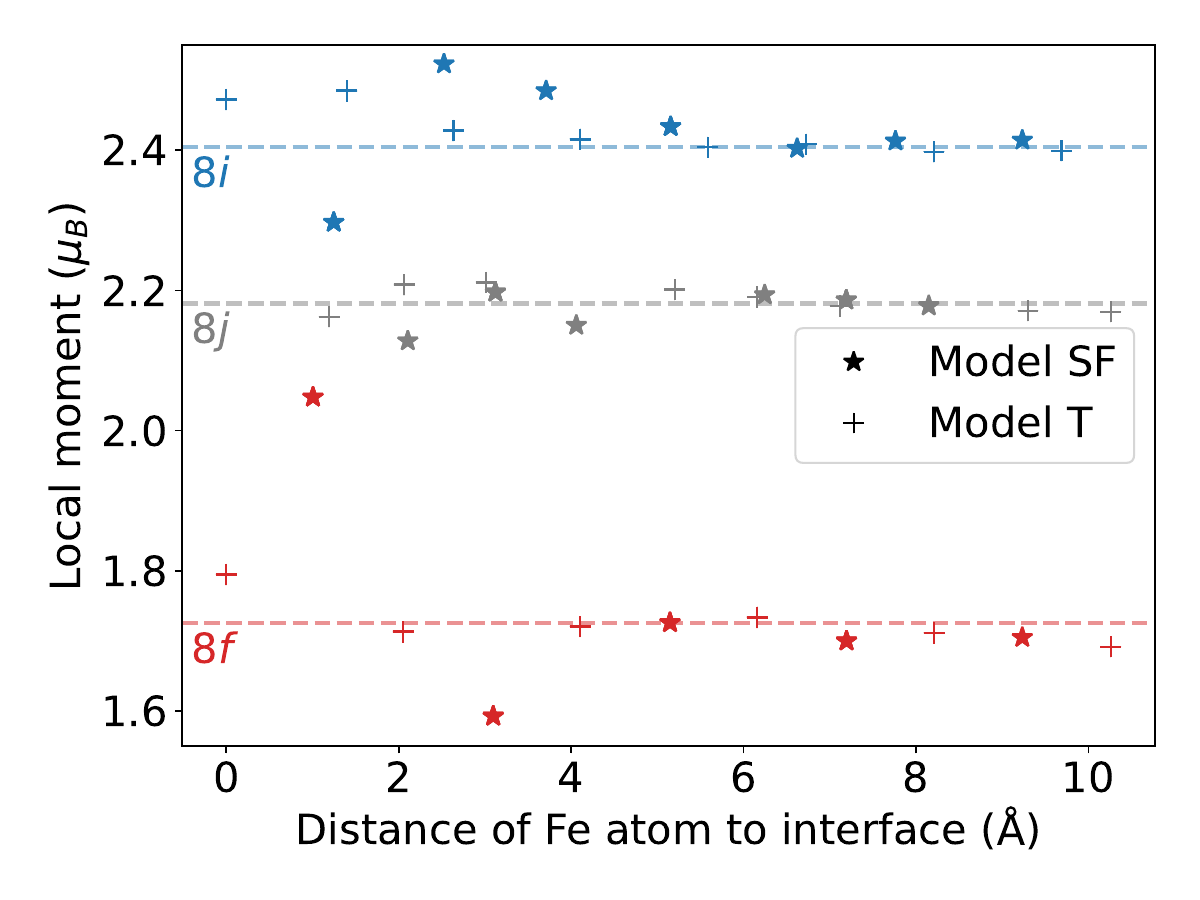}
    \caption{Local magnetic moments of Fe atoms for the
    interface models (symbols), 
    compared to bulk (horizontal dashed lines).
    Blue/gray/red corresponds to crystallographic $8i$/$8j$/$8f$
    positions.
    Note that zero distance corresponds to the reflection
    plane in Model T, while for model SF it is the plane
    directly between adjacent RT\textsubscript{4} planes
    (which contains no atoms).
    \label{fig.moments}
    }
\end{figure}

Figure~\ref{fig.moments} shows the magnetic moments of
the Fe atoms in the relaxed interface models.
The local moment size is most strongly affected by whether
the atom sits at the crystallographic 8$i$, 8$j$ or 8$f$
position.
For either model, all atoms at a distance greater than 4~\AA \ 
from the interface deviate by less than 0.03~$\mu_B$ from
their bulk values.
Indeed, for Model T, all local moments are within 0.1~$\mu_B$ of their bulk values, even the 8$i$ and 8$f$ atoms located
directly in the mirror plane.
Therefore, just as Fig.~\ref{fig.fig_RDF} emphasized the
similarity between Model T and the bulk in terms of atomistic
structure, Fig.~\ref{fig.moments} demonstrates this similarity
in terms of electronic structure.
However, in Model SF the Fe atoms that are closest to the interface
do show stronger deviations, particularly the four 8$f$ atoms
located 1.0~\AA\ from the interface.
These atoms have moments of 2.05~$\mu_B$, an enhancement
of 0.33~$\mu_B$ compared to bulk.
The local moments of the Y atoms
show much smaller deviations from their bulk value, 
of order 0.01~$\mu_B$.

\subsection{Crystal field coefficients}

In bulk YFe\textsubscript{12}, there is one unique
Y site, and its high symmetry leads
to nonzero crystal field (CF) coefficients 
$A_{lm}\langle r\rangle^l$
only for $(l,m)$ equal to $(2,0)$, $(4,0)$, $(4,\pm4)$
$(6,0)$ and $(6,\pm4)$.
However, the lower symmetry of the interface models
means that there are
three distinct Y sites in both Model T and Model SF,
as can be deduced by inspecting Fig.~\ref{fig.fig_atomistic}.
There are also no restrictions on
the $(l,m)$ values which result in nonzero CF coefficients.
Therefore, CF coefficients were calculated 
for all possible $m$ values when $l=2,4,6$.
It should be noted that CF coefficients depend on the 
coordinate frame used to calculate them.
Here, the necessary rotations were applied so that both models
and the bulk structure shared a common reference frame.

\begin{figure}
    \includegraphics[width=\columnwidth]{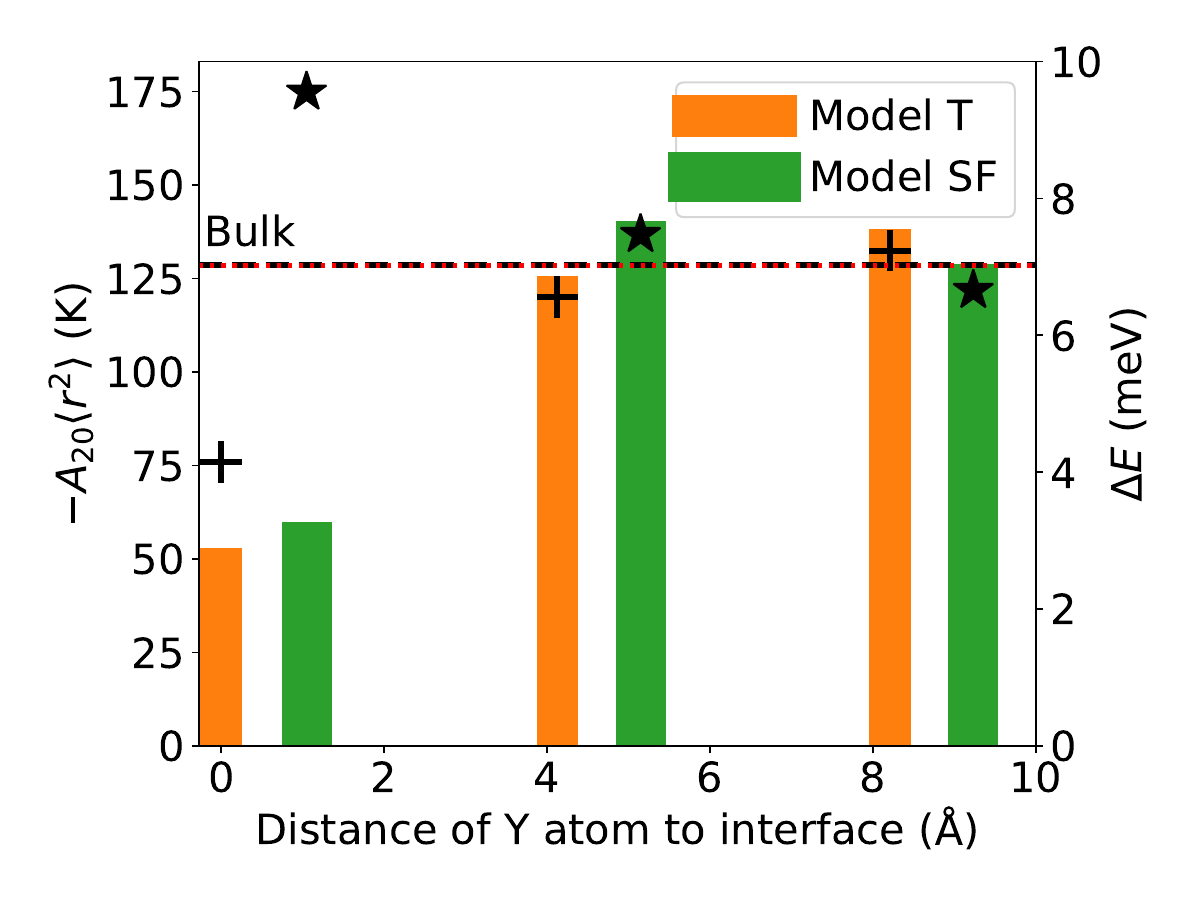}
    \caption{  $A_{20}\langle r^2\rangle$ crystal field 
    coefficients 
    (bars, left axis) and classical anisotropy energy 
    $\Delta E$ (symbols, 
    right axis) for the different interface models.
    Bulk values are shown as horizontal lines.
    \label{fig.A20}
    }
\end{figure}

Figure~\ref{fig.A20} shows the (negative) values of
$A_{20}\langle r\rangle^2$ for the different Y sites in
the models, as a function of distance from the
interface plane.
$-A_{20}\langle r\rangle^2$ is often the focus of analysis since,
under certain conditions and approximations, it
is proportional to the conventional anisotropy 
constant~\cite{Delange2017}.
The value of 129~K calculated for bulk YFe\textsubscript{12}
is shown as a dashed line in Fig.~\ref{fig.A20}.
The first important feature in Fig.~\ref{fig.A20} is that the
R sites located away from the interface plane, at a distance greater
than 4~\AA, have very similar CF values to bulk.
This is true for all $(l,m)$ values considered, not just $(2,0)$.
This mirrors behavior found previously~\cite{Patrick2024},
where the perturbation to the crystal field due to the introduction
of point defects had a limited range (again 4~\AA).

However, the magnitude of $A_{20}\langle r\rangle^2$ is significantly
reduced for the R sites closest to the interface plane.
It is important to realize that this does not necessarily 
mean that the single-ion anisotropy energy 
at these sites is smaller than in bulk, since this quantity is 
affected by all $(l,m)$ coefficients,
not just $(2,0)$.
However, the reduced magnitude of $A_{20}\langle r\rangle^2$ does show
that the magnetization is less inclined to point along the $c$-axis
direction.
This behavior is not surprising, particularly for Model T, where 
sites at the twin plane have two possible $c$-axes to choose from
(one for each grain), and there is no reason to favor one above the other.
Indeed, calculations below find that the easy direction at this site
points along the average of the $c$-axes of the individual grains.

\subsection{Classical anisotropy energy and easy directions}

The behavior of $A_{20}\langle r\rangle^2$ in 
Figure~\ref{fig.A20} and of the other CF coefficients 
(not shown) leads to the conclusion that
only the atoms in the immediate vicinity of
the interface structures in Fig.~\ref{fig.fig_models} 
have strongly-modified crystal fields.
In order to evaluate the effect on the anisotropy energy,
the calculated CF coefficients could be fed 
into the ``standard model'' spin Hamiltonian for
rare earth magnets 
to obtain the energy level splitting~\cite{Kuzmin2008}.
However, since this Hamiltonian also involves 
the exchange field generated by
the transition metal, it would be 
difficult to isolate the contribution
of the modified crystal field.
Instead, the CF coefficients are here used
to construct the classical single-ion anisotropy energy 
$E(\theta,\phi)$, through~\cite{Patrick2024}
\begin{eqnarray}
E(\theta,\phi) &=& \sum_{lm}\mathcal{A}_l
B_{lm} \ 
d_{l}^{0m}(\theta)
e^{im\phi}  \nonumber \\
&=&
\sqrt{\frac{4\pi}{2l +1}}
\sum_{lm}\mathcal{A}_l
B_{lm}
Y_{lm}(\theta,\phi)
\label{eq.class}
\end{eqnarray}
The angles $(\theta,\phi)$ describe the orientation of the 
local magnetic moment at that site.
The prefactor $\mathcal{A}_l$ depends on the total
angular momentum and Stevens 
coefficients~\cite{Stevens1952,Sievers1982},
and $d^{0m}_l(\theta)$ is related to the associated
Legendre polynomials, which with $e^{im\phi}$
can be expressed in terms of the complex spherical
harmonics~\cite{Edmonds}.

\begin{figure*}
    \includegraphics[width=1.7\columnwidth]{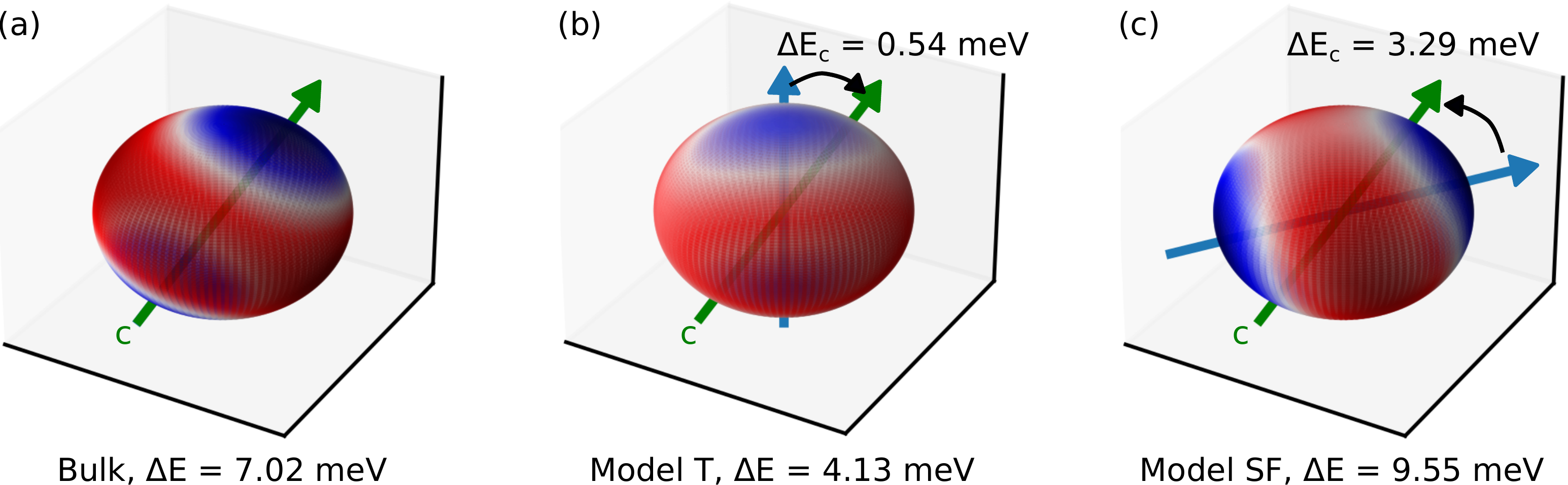}
    \caption{
    Energy as a function of moment orientation calculated
    using equation~\ref{eq.class}
    for bulk SmFe\textsubscript{12}
    or for the Sm site closest to the interface plane in the two
    models.
    In the color maps, blue and red correspond to the easiest
    and hardest directions, respectively.
    The blue and green arrows denote the local easy direction
    and $c$-axis, respectively (only one of the two possible 
    $c$-axes is shown for Model T).
    $\Delta E$ and $\Delta E_c$ are defined in the text.
    \label{fig.class}}
    \end{figure*}

Equation~\ref{eq.class} has been used to calculate the
single-ion anisotropy at all the R sites in the interface,
using $\mathcal{A}_l$ values appropriate for Sm.
Figure~\ref{fig.class} presents the results for sites located
closest to the interface plane in each model, and also 
for bulk SmFe\textsubscript{12}.
At each site one can find the local easy direction 
which minimizes $E$, and also define a local anisotropy energy
$\Delta E$ as the difference between the maximum and minimum 
of $E$.
It is also useful to introduce the quantity $\Delta E_c$, which 
is the energy cost of rotating the moment from its local 
easy direction to the direction parallel to the $c$-axis.

In bulk [Fig.~\ref{fig.class}(a)] the easy direction coincides
with the $c$-axis and there is a large $\Delta E$ value of 
7.02~meV (this value reduces if Ti is present at neighboring
8$i$ sites~\cite{Patrick2024}).
In the interface models, R sites at a distance greater
than 4~\AA \ from the interface plane have similarly large
values of $\Delta E$.
These values are shown as symbols in Fig.~\ref{fig.A20},
which shows how
the variations in $\Delta E$
for these sites follow the variations in 
$-A_{20}\langle r^2\rangle$.
Three of these four sites also have easy directions
along the $c$-axis,
but the site in Model SF located 5.1~\AA\ from the interface
has an easy direction which is rotated by $5^\circ$ with
respect to $c$.

The R sites closest to the interface planes show much stronger
variations in $\Delta E$ and have substantial differences
in easy direction.
For Model T, based on its symmetry 
the easy direction of the R site within the interface
plane should either point perpendicular
or parallel to this plane.
The calculated anisotropy energy [Fig.~\ref{fig.class}(b)]
confirms this, and shows that 
the easy axis is perpendicular to the interface plane,
pointing exactly halfway between the $c$-axes of the twinned
regions ($29^\circ= \tan^{-1}(c/a)$ with respect to $c$).
This direction can equivalently be seen as the average 
of the two grains' easy directions.
Furthermore, at this site $\Delta E$ is reduced to 4.13~meV, and
the energy cost $\Delta E_c$ of rotating the moment from this
easy direction to $c$ is only 0.54~meV.

In Model SF [Fig.~\ref{fig.class}(c)] the difference from
bulk is dramatic.  
The $\Delta E$ value of 9.55~meV 
is larger than in bulk, and the easy direction is strongly
misaligned from $c$, at an angle of 49$^\circ$.
$\Delta E_c$ is also substantial, 3.29~meV.
The behavior of this site shows the importance of considering
all CF coefficients, not just $A_{20}\langle r^2\rangle$; looking
only at this number (bars in Fig.~\ref{fig.A20})
might have led to the conclusion 
that the anisotropy at this site was weak.
It is also interesting to note the anisotropy in the plane
perpendicular to the easy direction, evident from the color map
in Fig.~\ref{fig.class}(c).
The RDFs in Fig.~\ref{fig.fig_RDF} help
explain the different single-ion anisotropy of 
this site.
The local environment (closer than 4~\AA, previously
found to be a critical radius for affecting CF 
coefficients~\cite{Patrick2024})  is substantially different
compared to bulk, with a new R site and reduced Fe coordination.
Given their +3 charge state, 
the R ions' positions are likely 
to be particularly important in determining
the orientation of Sm's prolate 4$f$ electron cloud.

\begin{figure}
    \includegraphics[width=\columnwidth]{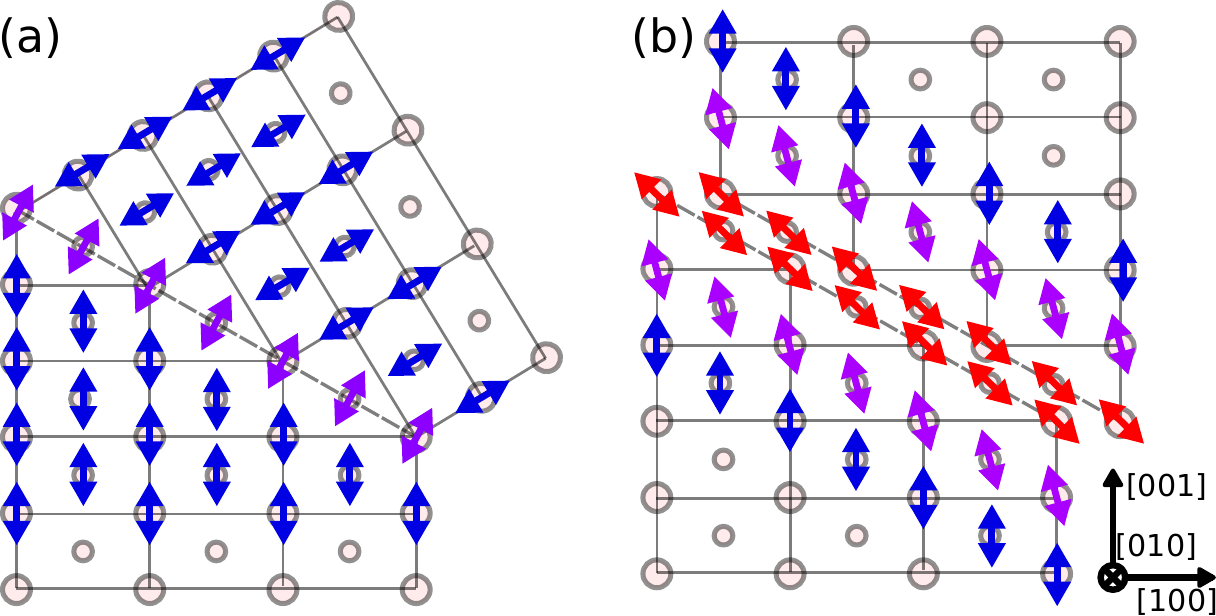}
    \caption{   Schematic representation 
    (see also Fig.~\ref{fig.fig_construction}) of the local easy
    axes at different R sites in (a) Model T and (b) Model SF.
    All easy axes lie in the plane of the page $(010)$.  The blue
    arrows are parallel to the $c$ axis.
    \label{fig.easydirections}
    }
\end{figure}

Figure~\ref{fig.easydirections} summarizes the calculations
of the single-ion anisotropy, showing the easy axes 
at the different sites explored in the calculations.
All of the easy axes lie parallel to the $(010)$ plane.
These diagrams emphasize how the additional RT\textsubscript{4}
plane in Model SF causes a major disruption in anisotropy,
creating a region approximately 2~\AA \ wide where the 
easy direction
is $49^\circ$ to the $c$ axis, and a wider region (approximately
10~\AA) where the influence is much weaker, but
still noticeable as a small change in easy direction.
By contrast, the mirror plane in Model T shows relatively simple
behavior, with an easy axis which is an average of the $c$-axes
of the two regions.
It should be noted that this direction would also 
correspond to the global easy axis of a crystal 
composed of two identical
grains twinned in this way, since this is determined by
the same average.

\section{Conclusions and outlook}
\label{sec.conclusions}

Recently published HAADF-STEM images have been used
to derive atomistic models of the two reported twinning
structures.
One structure is a simple reflection in a \{101\} plane,
which has no effect on the stoichiometry. 
The R atoms at this interface have essentially
identical RDFs to bulk RT\textsubscript{12}.
The other structure involves the insertion of an RT\textsubscript{4}
plane followed by an offset of $a/2$ in the [100] direction 
(loosely described as a stacking fault)
followed by a reflection.
This structure is R-rich and the RDFs show
large differences compared to bulk.
Two periodically-repeating models have been constructed,
Model T and Model SF, to investigate the reflection
and the stacking fault respectively, and their
properties have been calculated using first-principles DFT.
Analysis both of magnetic moments and crystal field coefficients
finds that only the atoms within $\sim$4~\AA \ of the interface
plane are strongly affected by the reflection or stacking fault.
In Model T, the R atoms in the reflection plane show a weakened
single-ion anisotropy and have an easy direction equal to 
the average of the $c$-axes of the two grains.
However, in Model SF the R atoms closest to the interface have
an enhanced single-ion anisotropy compared to bulk, with an 
easy direction $49^\circ$ to the $c$-axis.
The effects on the next-closest R atoms are much smaller, but
the easy direction is slightly rotated away from $c$.

This work has addressed the first key question in 
Section~\ref{sec.intro}, and quantified the effect
of experimentally-observed extended defect structures on
the microscopic anisotropy.
The ``stacking fault'' does indeed induce substantial 
changes in single-ion anisotropy, but only in the atoms
closest to the interface.
The pressing issue now is address the second key question
and understand what effect these
microscopic changes have on the macroscopic phenomena of
magnetization reversal and coercivity.
Although micromagnetics calculations may provide some 
indication, we note that the small length scale (2~\AA)
may not be well described by a continuum approximation.
Atomistic spin dynamics calculations do not make this approximation,
and have been used to study magnetization and 
domain wall structure in 
Nd\textsubscript{2}Fe\textsubscript{14}B~\cite{Toga2018,
Gong2019,Nishino20212}.
Therefore, the likely next step is to use the calculated
results as input for ASD simulations.

As noted in Section~\ref{sec.methods}, this study has 
focused on the simplest (and thermodynamically unstable)
RFe\textsubscript{12} composition, 
and not attempted to investigate substitutions of Co, Ti
or V.
In bulk, there is a strong tendency for Ti or V atoms to occupy
8$i$ sites, and as noted above, the RT\textsubscript{4} plane
contains two 8$i$ sites.
On the other hand, whilst the STEM-EDS measurements in
Ref.~\cite{Tozman2025} are not conclusive, there is 
some evidence for a depletion of Ti in the interface 
region~\cite{Tozman2025}.
It will be important to investigate the site preference energetics
of substitutional T atoms in these interface structures, and
also to confirm that their effects on the crystal field are in
line with those predicted by a point charge 
model~\cite{Patrick2024}.
More generally, first-principles calculations should be used 
to address the key question of understanding the relative stability
of the different types of interfaces.
If these different interfaces do indeed affect the coercivity, 
determining the conditions which promote the formation of one
over the other would be a major step forward in controlling
the performance of practical one-twelve magnets.
Finally, both micromagnetics and ASD
require knowledge of the exchange interactions.
Based on the relatively small changes in magnetic moments that 
were calculated for the interface models, it is likely that
bulk calculations could be used to parametrize the exchange
interactions, as a first approximation.
However, a first-principles calculation of exchange
coupling in the interface models introduced here 
would test the validity
of this approximation, and also provide further fundamental 
insight into the nature
of exchange across interfaces.

\begin{acknowledgments}
I am grateful for computational support from the UK national high performance computing service, ARCHER2, for which access was obtained via the UKCP consortium and funded by EPSRC grant ref EP/X035891/1.
I thank J.\ B.\ Staunton for a critical reading of the manuscript.
I also thank A.\ Bolyachkin and H.\ Sepehri-Amin for helpful
discussions about the data of Ref.~\cite{Tozman2025},
and A.\ Gabay, G.\ Hadjipanayis and L.\ H.\ Lewis for
initial discussions about one-twelve magnets.
\end{acknowledgments}


%

\end{document}